\documentclass[10pt]{iopart} \usepackage{graphicx,epsfig}

\begin{document}
\title[Dark solitons in atomic condensates and optical systems]
{Analogies between dark solitons in atomic Bose-Einstein condensates and optical systems}

\author{N.P. Proukakis$^{\dag}$, N.G. Parker$^{\dag}$, D.J. Frantzeskakis$^{\ddag}$, and C.S. Adams$^{\dag}$}
\address{$^\dag$ Department of Physics, University of Durham, South
Road, Durham DH1 3LE, United Kingdom}
\address{$^\ddag$ Department of Physics, University of Athens, Panepistimiopolis, Zografos,
Athens 15784, Greece}

\ead{n.p.proukakis@durham.ac.uk}

\begin{abstract}

Dark solitons have been observed in optical systems (optical fibers, dielectric guides and bulk media), 
and, more recently, in
harmonically confined atomic Bose-Einstein condensates. This paper presents an overview of some
of the common features and analogies experienced by these two intrinsically nonlinear systems, with emphasis
on the stability of dark solitons in such systems and their decay via emission of radiation.
The closely related issue of vortex dynamics in such systems is also briefly discussed.

 \end{abstract}
\maketitle

\section{Introduction}

Nonlinear systems in physics support the appearance of solitary waves \cite{dodd},
which propagate without dispersion. Such waves have been
extensively studied, both theoretically and experimentally in diverse
systems, such as shallow water waves \cite{water}, macromolecules
 \cite{macro}, acoustics \cite{acoustics}, plasma physics \cite{plasma},
elastic surfaces \cite{elastic}, optical fibers \cite{optical}, condensed excitons in crystals
\cite{exciton} and, more recently, realised experimentally in
atomic Bose-Einstein condensates (BECs) \cite{dark,bright}.
Solitary waves are of two kinds, depending on the sign of the effective nonlinearity,
whose physical origin and interpretation is system-dependent.
If the effective nonlinearity is attractive, then
bright solitons are formed, whereas dark solitons arise in the opposite case (e.g. defocusing media).
Bright solitons are non-dispersive (positive) density waves, whereas dark solitons correspond to
density depressions characterised by a phase shift across their density minimum. The behaviour of
these two distinct non-dispersive excitations is very different.

This paper focuses on the case of dark solitons. Dark solitons have been
investigated at length in nonlinear optics \cite{kivshar1}, where potential
applications in optical communications and in photonics
have also been proposed (see, e.g., relevant experimental results
in \cite{application1,application2}, and also \cite{application3}). This paper aims to highlight some
common features between optical and atomic dark solitons \cite{analogies}, in the context of
the dynamics of dark solitons in BECs. Although very different in origin,
the underlying equation describing the dynamical properties of dark
solitons in nonlinear optical systems and atomic
BECs has the same structure, given by the following dimensionless
Nonlinear Schroedinger Equation (NLSE) (usually referred to as Gross-Pitaevskii equation (GPE)
in the context of BECs)
\cite{review}
\begin{eqnarray}
i\frac{\partial \psi}{\partial t}=-\frac{1}{2} \nabla_{d}^{2}\psi+
F(|\psi|^2) \psi+V_{\rm{ext}}\psi.
\end{eqnarray}
In nonlinear optics, $\psi$ is the (complex)
electric field envelope, the variable $t$ plays the role of the
propagation distance $z$ (along the waveguide or fiber), while the
right-hand side has a different sense for temporal solitons
(in optical fibers) or spatial solitons (in bulk media or
dielectric waveguides). In the case of temporal solitons, $d=1$
and $\nabla_{1}^{2}\psi=\partial^{2} \psi/\partial \tau^{2}$
describes the normal dispersion (where $\tau$ is a retarded time
measured in a frame of reference moving with the group velocity),
while, in the case of spatial solitons, $\nabla_{d}^{2}\psi$
describes the beam's diffraction (here $d=2$ and $\nabla_{2}^{2}$ is
the transverse Laplacian). On the other hand, $F(|\psi|^2)$ is
proportional to the intensity-dependent change of the refractive
index of the optical medium
($I=|\psi|^2$ is the light intensity). In optical fibers, the nonlinearity is of the Kerr
type, i.e., $F(|\psi|^2)=|\psi|^2$. In bulk nonlinear media (such
as vapors, semiconductors, polymers, etc), the nonlinearity may be
of a non-Kerr type, e.g., a competing, or, generally, a saturable
nonlinearity \cite{kivshar1}, which, in some cases (e.g.,
relatively low light-intensities), can be approximated by the Kerr
nonlinearity. Finally, in the case of dark spatial solitons in
dielectric waveguides, the term $V_{\rm{ext}}$ accounts for a
possible change of the linear part of the refractive index in the
transverse dimensions. As an example, in an inhomogeneous
nonlinear waveguide, the refractive index $n_{w}$ may take the form
$n_{w}=n_{0}-n_{1}(x^{2}+y^{2})-n_{2}|{\bf \psi}|^{2}$, where
$n_{j}>0$ ($j=0,1,2$) represent the homogeneous, inhomogeneous and
nonlinear (Kerr) parts, respectively (see, e.g., \cite{agra}). This case bears
resemblance to the typical situation occuring in BECs, where the
condensate is usually trapped in a parabolic magnetic potential
\cite{review} (see below). Notice that in the case of the
nonlinear waveguide, the propagation distance $t$ in equation (1) is
measured in units of the diffraction length
$L_{D}=\beta_{0}w_{0}^{2}$ ($\beta_{0}$ is the longitudinal
wavenumber and $w_{0}=(2\beta_{0}^{2}n_{1}/n_{0})^{-1/4}$ is a
transverse length scale), the transverse coordinates $x$, $y$ in
units of $w_{0}$ and the electric field envelope in units of
$(\beta_{0}L_{D}n_{2}/n_{0})^{-1/2}$.

In the context of BECs, $\psi$ corresponds to the macroscopic
order parameter of the system at zero temperature (or, in general.
for temperatures much lower than the temperature at which the BEC
phase transition occurs), and this can be thought of as the wavefunction
describing the condensate. It is a complex parameter that can be
expressed as $\psi=\sqrt{n}\exp(i\phi)$, where $n$ is the atomic
density and $\phi$ its phase.  The terms appearing on the
right hand side of equation (1) denote, respectively, the kinetic
energy contribution in a $d$-dimensional manifold (in the general
case, $\nabla_{d}^{2}$ with $d=3$ is the transverse Laplacian in
$3$-dimensions), the strength and form of the nonlinearity and the
external confinement of the system under consideration. It should
be noted that in atomic BECs, the nonlinearity is also an
intrinsic phenomenon, arising from the scattering properties
between the atoms of the condensate. Nevertheless, for
sufficiently dilute atomic gases, only s-wave two-body scattering
is important, and, as a result the corresponding nonlinear term in
equation (1) can usually be approximated as $F(|\psi|^2)=|\psi|^2$,
resembling the Kerr nonlinearity term appearing in
optics. The last term can generally account for any external
confinement imposed on the medium in which soliton propagation may
take place. Presence of such a spatially-dependent term
(accounting for the graded-index waveguide in optics or the
confining potential in BECs) modifies the local background density
in the medium, which thus becomes non-uniform. Excluding
spatially-dependent refractive indices, optical waveguides are
generally homogeneous systems, whereas atomic BECs (unlike other
superfluids, such as $^{4}$He) are typically formed in axially symmetric harmonic
(magnetic) traps, featuring inhomogeneous confinement along all
three of the system's directions, with  $V_{\rm ext}=(1/2)m(\omega_z^2
z^2+\omega_r^2 (x^2+y^2))$, where $\omega_{z}$ ($\omega_{r}$) the longitudinal
(transverse) confining frequency.
This leads to very distinct
dynamics between the two systems, since dark solitons in BECs
propagate on a position- (and time-) dependent background, in stark
contrast to the constant background in optical waveguides

This paper focuses on the case of a cubic nonlinearity, which is relevant to both
optical and atomic systems. In this
case, strictly speaking (and from a mathematical point of view), a
NLSE dark soliton is a solution of equation (1) in one dimension
(i.e. $d=1$, for which $\nabla_{1}^{2}=(\partial^{2}/ \partial
z^{2})$), on a homogeneous background density (i.e. in the absence
of the third term appearing above), which reduces to
\begin{eqnarray}
i\frac{\partial \psi}{\partial
t}=-\frac{1}{2}\frac{\partial^{2}}{\partial z^{2}}\psi+|\psi|^2
\psi.
\end{eqnarray}

In order to reduce the 3D Gross-Pitaevskii equation to 
the above 1D form, one requires (i) very tight transverse (radial) confinement,
such that transverse excitations are completely suppressed, and 
(ii) a longitudinally homogeneous density ($\omega_{z}=0$).
In situations of tight
radial confinement, one obtains an  effective
one-dimensional interaction strength $g$
by integrating the three dimensional interaction strength $g_{3D}$
over the transverse directions.
Since $g_{3D}=4\pi \hbar^2 a/m$ (where $a$
is the {\it s}-wave scattering length characterising the atomic
interactions and $m$ the atomic mass), this yields
$g=g_{3D}/(2\pi l_r ^2)$, where
$l_r=\sqrt{\hbar/m\omega_r}$ is
the transverse harmonic oscillator length.
To further reduce the resulting equation to dimensionless form,
length is scaled in units of the fluid healing
length $\xi=\hbar/\sqrt{n_0g m}$, velocity in terms of the
Bogoliubov speed of sound for the medium $c=\sqrt{n_0g/m}$, and
the atomic density rescaled by the peak density $n_0$.  Energy is
scaled in terms of the chemical potential of the
system $\mu=gn_0$. 

The wavefunction of a dark soliton propagating with speed $v$ and
position $(z-vt)$ on a uniform background of unity is given
analytically by
 \cite{sol}
\begin{eqnarray}
\psi(z,t)=e^{-it} \left( \lambda \tanh \left[ \lambda
\left(z-vt\right)\right]+i v \right).
\end{eqnarray}
Here $\lambda=\sqrt{1-(v/c)^2}$, and the healing length $\xi$
corresponds roughly to the size of the soliton. The soliton speed
$v$ depends on its depth $n_d$ relative to the background density
and the phase slip $S$ across its centre via
$v/c=\sqrt{1-(n_d/n_0)}=\cos (S/2)$.

The fact that dark solitons are characterized by a nontrivial distribution of their phase 
(actually, initially, i.e. at $t=0$, the soliton phase is an odd function of $x$), raised certain technical difficulties for the experimental verification of their propagation in optical systems. That is why the first experimental attempts to study dark temporal solitons in optical fibers took place in the late $1980$'s, 
even though dark solitons were predicted to occur in optical fibers as early as 1973 \cite{hasegawa}. 
These first experiments reported the creation of a 
fundamental dark soliton [i.e., $\upsilon=0$ in equation (3)] (using a $\pi$-phase step) 
\cite{emplit}, the evolution of a pair of small-amplitude dark solitons (emerging from an even dark pulse) \cite{krokel}, as well as the generation and subsequent evolution of a dark soliton  (emerging from a background pulse with a phase jump) \cite{weiner}. Later, during the $1990$'s, the generation of dark soliton trains at high repetition rates \cite{trains}, as well as their potential applications in optical communications [12] was demonstrated. Note that the first experimental results for the generation of spatial dark solitons (upon using proper amplitude and phase masks) were reported in sodium vapor 
\cite{swartz}, bulk semiconductors \cite{allan}, photorefractive materials \cite{duree}, etc., at
around the same time.

On the other hand, matter-wave dark solitons have also been observed in several recent experiments [9]. Similarly to the case of the optical systems, a quantum-phase engineering technique (or phase-imprinting method) was used to imprint the appropriate phase distribution on the BEC so as to create dark solitons. The phase-imprinting method (which was also employed for the creation of vortices in BECs \cite{Vortex_Imp,murray}) is an efficient tool to engineer the phase of the BEC clouds \cite{dobrek,bongs2,Dum} and its optimization has been discussed in \cite{carr}. In particular, in that work it was suggested that such an optimization involves engineering not only the phase, but also the BEC density, which may result in the generation of stationary dark matter-wave solitons. Apart from the phase-engineering technique and the above mentioned optimized version of it (i.e., the so-called phase and density engineering technique \cite{carr}), Burger et al. \cite{carr2} recently proposed the creation of dark solitons in BECs by {\it purely} engineering the density distribution. Such a ``density engineering'' technique bears resemblance to the creation of optical solitons by intensity modulations of a backround light field, as in relevant experiments in optical fibers (see, e.g., \cite{krokel} and relevant theoretical work in \cite{gred}).

The idealised soliton of equation (3) propagates in a stable manner. However,
realistic conditions deviate from this idealised equation
(equation~(2)), thus breaking the integrability of the system and
rendering the soliton unstable to decay via sound emission. There
are different types of instabilities which can arise. Firstly, a
dark soliton in usually embedded in higher than one-dimensional
(1D) geometry. This leads to additional kinetic energy
contributions in the transverse directions, which are expected to
lead to the dominant decay mechanism, if the system is far from
the 1D limit. Optical fibers are essentially 1D objects, thus
eliminating, to large degree  such instabilities. Initial
experiments with condensates focused on 3D geometries. In lower
dimensionalities, long wavelength excitations tend to destroy the
universal phase coherence across the trapped atomic system (in
accordance with the disappearance of BEC in dimensions $d \le 2$
for a homogeneous system). However, it has recently been shown
that, at ultralow temperatures, the universal coherence across the
system is essentially restored \cite{gora,lowd}. This regime is
now a topic of active experimental and theoretical research.
Quasi-one-dimensional atomic systems can be realised by the
application of very tight transverse confinement \cite{gorlitz},
such that the transverse degrees of freedom become essentially
frozen out \cite{olshanii}. Such quasi-1D systems are particularly
important in so-called atom chips \cite{hansel,schreck,ott,ed},
where the motion is confined along only one direction by a current
carrying wire. Hence, the domain of quasi-1D is now also
accessible experimentally in atomic BECs. Experiments with dark
solitons have so far been performed in 3D BECs \cite{dark}, but
one can now examine their dynamics in quasi-1D systems \cite{kavoulakis}, in which
their stability is expected to be largely enhanced
\cite{muryshev}.

A second source of instability arises from modifications of the
above nonlinearity. In the general case of a non-cubic
nonlinearity, equation~(2) ceases to be a completely integrable
system and as a result dark solitons are subject to perturbations
that may be quite strong. This is particularly important,
especially in the case of optical media, where deviations from the
rather simple model of the Kerr nonlinearity may be significant.
In such cases, more realistic models, such as the saturable
nonlinearity, are relevant. Since the nonlinearity in atomic BECs
arises from interactions within the condensate atoms, an
additional dissipative mechanism will arise due to the presence of
thermal (uncondensed) atoms in the medium. This effect is
analogous to damping of the superfluid component in liquid helium,
due to coupling with the normal component \cite{landau}, and can
be suppressed by going to extremely low temperatures in the atomic
traps \cite{fedichev}. One should also mention that if the
confinement becomes extremely tight, the nonlinearity in atomic
BECs start deviating from the cubic form discussed throughout this
work \cite{kolomeisky}, and soliton propagation in this regime
will be discussed elsewhere. 
Additional decay mechanisms arise from modifications from a
homogeneous background, an important effect in atomic BECs.
For completeness, one should also
mention the effect of quantum fluctuations which have been shown
to lead to decay in the case of deep slow solitons
\cite{polish}. 

Soliton decay due to the above instabilities is typically accompanied by the
emission of sound waves.
These sound waves are density waves in atomic
condensates, and correspond  to electromagnetic radiation in the
optical context.
Note, however, that, under appropriate conditions, the soliton may
re-interact with the emitted sound, with the effect of {\em stabilising} 
it against decay. This occurs, for example, for
 dark solitons in quasi-1D condensates featuring longitudinal confinement,
where stabilisation arises as a result of 
continuous sound emission and reabsorption cycles \cite{parker1}.

This paper is organised as follows: Section 2 discusses
instabilities due to transverse excitations, whereas the rest of
the paper focuses on the limit of tight transverse confinement,
where this effect is negligible. In section 3 we highlight
analogies between dark solitons propagating through potential
steps and soliton transmission through regions of different
indices of refraction. Section 4 outlines the dominant decay
mechanisms in one-dimensional systems arising in both atomic and
optical systems. A common result for the sound emission is found
to be obeyed in this limit. In section 5, we briefly address the
issue of stabilisation of dark solitons against decay mechanisms.
Section 6 discusses analogous effects in vortices, a higher
manifold topological structure which has recently been observed in
both optical and atomic systems. Our concluding remarks are
presented in section 7.

\section{Transverse Instabilities}

\subsection{Optical Dark Solitons}

In optical systems a tight transverse confinement of light beams
can be realized in waveguide geometries (such as the optical fiber
or the slab waveguide), in which the dark solitons are practically
1D objects, that are not prone to transverse instabilities.
However, in planar waveguides or in bulk media, where the optical
systems become effectively 2D or 3D respectively, the rectilinear
dark solitons (i.e., the dark soliton stripes) are subject to a
long-wavelength transverse instability (the so-called ``snake
instability''), which leads to the stripe breakup and the eventual
creation of optical vortex solitons with alternate topological
charges. The instability band is characterized by a maximum
modulation wavenumber $Q_{cr}$ (depending on the dark-soliton
amplitude), i.e., the dark soliton stripe is stable only for
$Q>Q_{cr}$, where $Q$ is the perturbation wavenumber \cite{kuz}.

In the early experiments \cite{swartz,ee}, where dark soliton stripes and grids
were created on the transverse cross-section of an optical beam propagating
through a bulk self defocusing medium (i.e., sodium vapor), the snake instability
was not observed due to the finite-size of the beam and the weak
nonlinearity of the medium. However, in later experiments in rubidium vapor
\cite{tikh} (relevant results have also been reported using photorefractive
crystals \cite{snake_opt}), the enhancement of the nonlinearity of the
optical medium with increasing temperature, led to the direct
experimental observation of the snake instability of a dark soliton stripe
and the subsequent creation of vortex solitons (see also \cite{instability}
for a review).

\subsection{Dark Solitons in Atomic BECs}

The snake instability arises in atomic systems as well; there,
recent experiments have demonstrated the decay of the rectilinear
dark solitons into vortex rings \cite{snake_BEC}. As in the case
of nonlinear optics, this decay mechanism dominates far from 1D
geometries, with the longitudinal sound emission being negligible
in comparison. The effect of the snake instability was studied
theoretically in a series of works \cite{s_th_1,s_th_2} and it has
been suggested that it can be suppressed by a sufficiently strong
transverse confinement of the condensate \cite{muryshev}, which
bears resemblance to the use of finite-size beams in nonlinear
optics. Another possibility for the suppression of the snake
instability is to bend a dark soliton stripe to form a ring of
length $L<2\pi/Q_{cr}$. Such ``ring dark solitons'' were predicted
theoretically \cite{kyang} and observed experimentally
\cite{neshev} in the context of optics. Recently, ring dark
solitons were predicted to occur in BECs as well \cite{gt}, with
the snake instability leading to robust vortex arrays in the form
of necklaces. Suppression of the snake instability in BEC
experiments can be achieved, for example, in highly-elongated
(``cigar-shaped'') geometries. Here we aim to study the mechanism
of the snake instability in BECs, upon considering a
cylindrically-symmetric 3D geometry, explicitly featuring
longitudinal harmonic confinement.

\begin{figure}
\begin{center}
\includegraphics[width=10.0cm]{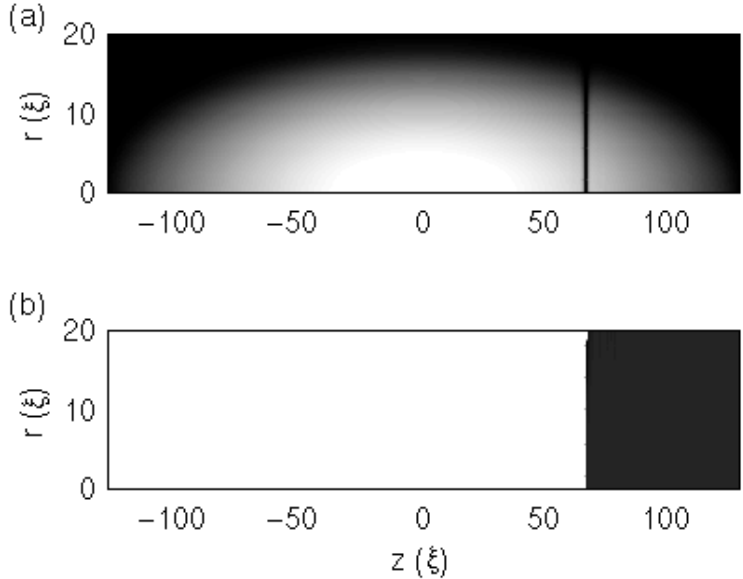}
\caption{Planar view of (a) density and (b) phase profiles of a
stationary off-centred dark soliton in an atomic Bose-Einstein
condensate confined in a cylindrically-symmetric 3D trap. Here z
denotes the longitudinal and r the transverse direction. The dark
soliton, positioned at $z=66\xi$, appears as a notch of zero
density in the z-r plane and clearly features an abrupt phase slip
of $\pi$. The trap frequencies are $\omega_z=\sqrt{2} \times
10^{-2}$ $(\mu/\hbar)$ and $\omega_r=0.1$ $(\mu/\hbar)$, giving a
trap ratio $\omega_r/\omega_z \sim7.1$. The chemical potential of the
system, $\mu$, in terms of which energies are scaled, is obtained,
by fixing the peak 3D density of the system to 1.  
Note that, in the density (phase) scale, white
represents the peak density $(\pi/2)$ and black represents zero
$(-\pi/2)$.}
\end{center}
\end{figure}
\begin{figure}
\begin{center}
\includegraphics[width=10.0cm]{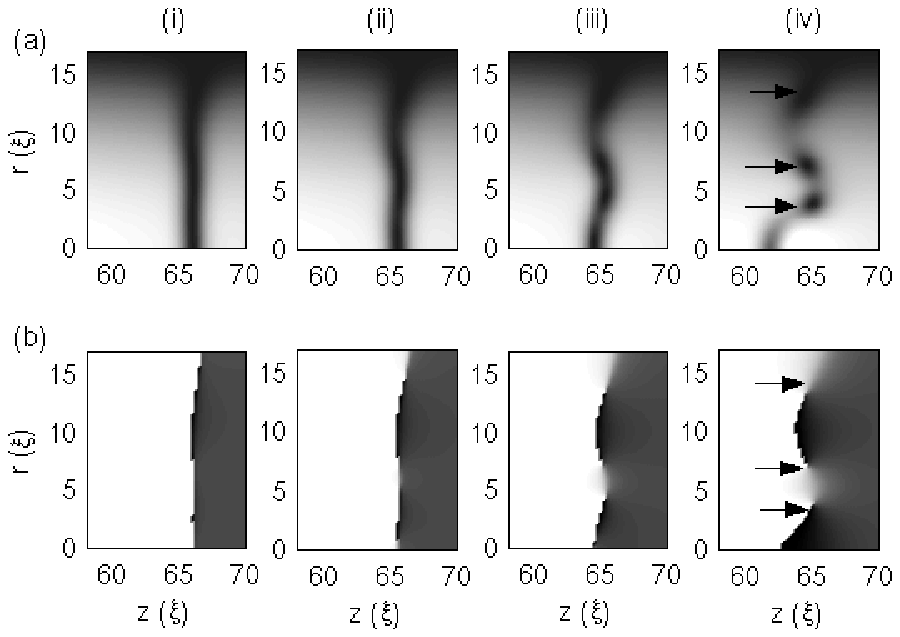}
\caption{Close-up snapshots of the evolution of (a) density and
(b) phase profiles for the initial soliton configuration of
figure~1, depicting the main stages in the dynamics of the snake
instability. Successive images correspond to $t\approx$ (i) $15$,
(ii) $19$, (iii) $23$ and (iv) $27$ $\xi/c$. In (iv), the
dark soliton has already decayed into 3 vortex rings, as evident
by the corresponding phase profile. The location of these vortex
rings is indicated by arrows.}
\end{center}
\end{figure}

The transverse stability of a dark soliton in a harmonic trap can
be studied by monitoring the evolution of a stationary dark
soliton created at an off-center position.  
This is based on numerical simulations of the cylindrically-symmetric GPE.
The initial density and phase profiles in the $z$-$r$
plane are shown in figure~1, (where $z$ is the longitudinal and
$r$ the transverse direction). The phase profile (b) features a
step-like $\pi$ phase slip across the soliton minimum. Due to the
inhomogeneous background density, the soliton accelerates towards
the center, with its motion being subject to strong coupling between the
longitudinal and transverse degrees of freedom. Since the soliton
is not the lowest energy state in 3D geometries, it tends to decay
into more stable, lower energy structures, such as vortex rings.
This occurs due to the existence of a mode with imaginary
frequency components
 \cite{s_th_1,s_th_2}. To probe this decay mechanism, figure~2 shows
successive (a) density and (b) phase snapshots of a soliton in a
three-dimensional geometry, focusing on the interesting regions
around the soliton minimum. One clearly observes the anticipated
snake-like bending of the soliton plane, followed by the decay
into vortex rings and sound waves (the latter not easily visible on the greyscale
used in this subsection, as their density is typically much less than
that of the vortex rings). Cross-sections of these rings in the $z$-$r$ plane
can be seen in (iv), with each vortex ring indicated by an arrow.

Such a decay mechanism involving the bending of the soliton plane
will clearly tend to be suppressed as the transverse size of the
condensate decreases, leading eventually to its complete
prevention in highly-elongated quasi-1D geometries. The regimes of
transverse stability of a moving soliton have been studied in
\cite{muryshev1}. To highlight the effect of transverse
confinement on soliton stability,  figure~3 shows density and
phase profiles of the evolution of an initially off-centred
stationary soliton at the time when the soliton (or its decay
products) have reached the trap centre, in traps of different
transverse confinement. The top figure corresponds to the unstable
3D case examined in figures 1 and 2. In this case, the $2\pi$
vortex phase singularity exhibited by the central vortex ring is
evident in figure 3(b)(i), having been highlighted by a hollow
white circle. Tightening the transverse confinement leads to a
decrease in the bending of the soliton, and hence the production
of less vortex rings, an issue studied in detail in
\cite{komineas}. Figure 3(a)(ii) shows the case of a single vortex
ring being produced, with the corresponding phase profile labelled
in figure 3(b)(ii). Finally, very tight transverse confinement,
leads to an effectively 1D system, featuring a stable soliton, as
evident from the step-like phase profile of figure
3(b)(iii).

\begin{figure}
\begin{center}
\includegraphics[width=8.0cm]{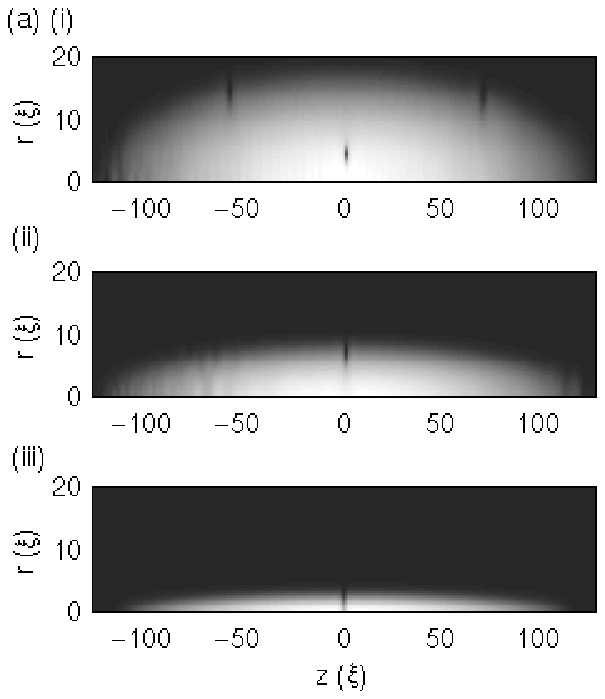}\includegraphics[width=3.0cm]{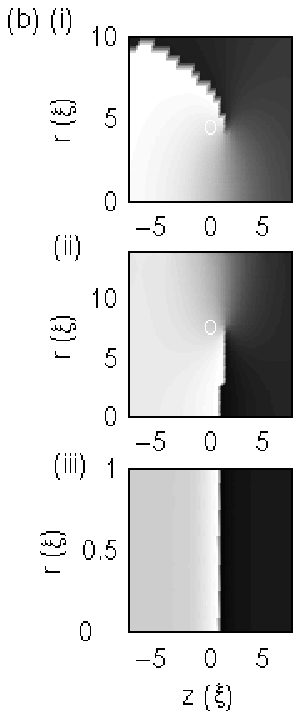}
\caption{(a) Density (left) and (b) phase (right) profiles of the
evolution of an initially off-centred stationary dark soliton for
different transverse confinement. All simulations start with a
dark soliton located at $z=66\xi$, with the images shown at $t=$
(i) $279$, (ii) $193$, and (iii) $160$ $\xi/c$, after the soliton
creation. From top to bottom, the aspect ratio of transverse to
longitudinal harmonic confinement is ($\omega_{r}/\omega_{z}$)$=$
(i) $7.1$, (ii) $14.2$, and (iii) $35.5$, 
corresponding to ($\mu/\hbar \omega_{r}$) = (i) $10$, (ii) $5$, (iii) $2$,
for the employed longitudinal
trap frequency $\omega_z=\sqrt{2} \times 10^{-2}$ ($\mu/\hbar$).
(i) Dark soliton has already decayed into 3 vortex rings, with the
$2 \pi$ phase singularity of the central vortex ring shown in
(b)(i), and highlighted by the hollow white circle. (ii) Soliton
decays into a single vortex ring (phase profile on the right).
(iii) Soliton in very elongated quasi-1D condensate is stable
against the snake instability, as shown by the step-like $\pi$
phase difference across it. }
\end{center}
\end{figure}

Lifetimes of dark solitons are hence clearly enhanced in
geometries featuring tight transverse confinement, for which the
chemical potential is too small to allow for transverse modes to be excited
due to atom-atom interactions (or thermal effects).
Any further studies of dark
soliton dynamics will therefore be performed in highly elongated
geometries, where the above decay mechanism is largely suppressed,
and this will be assumed in all subsequent sections. All following
results are therefore obtained by solving the 1D GPE with
a suitable potential $V(z)$.

\section{Propagation through regions of different potential energy}

\subsection{Optical Solitons}

One interesting application of nonlinear optics is in nonlinear waveguides,
which have been proposed as all-optical devices that may be used as switchers,
modulators, bistable-optical elements, etc. An important effect in this
context is the self-localization of optical beams, which leads to the
formation of finite-size self-focused channels, as well as to novel
stationary nonlinear guides and surface waves in thin-film
planar waveguides and at dielectric interfaces (see, e.g., \cite{mn}).

Considering that the self-focused light channels
can be well-approximated by spatial bright solitons, it has been shown
\cite{aceves} that their dynamics near interfaces separating
linear and nonlinear media or different nonlinear media, can be reduced to
the study of motion of an equivalent particle in an effective step-like
potential. This so-called ``equivalent particle theory actually'' corresponds
to the adiabatic (leading-order) approximation of the perturbation theory
of solitons \cite{yuriboris}, while a higher-order approximations allows for
inclusion of radiation effects following the beam reflection \cite{kivshar2}.
Adopting this approach, the propagation of light beams in planar thin-film
nonlinear waveguides has been studied in detail \cite{kqt}.

To link the above discussion with the following one referring to
BECs, it is important to stress that in all the above works, as
well as in earlier relevant contributions
\cite{kivshar3,kivshar4}, the scattering of an optical beam by an
interface separating two media of different refractive indices can
effectively be described by the dynamics of a NLSE soliton in a
step-like potential.

\subsection{Solitons in Atomic BECs}

\begin{figure}
\begin{center}
\includegraphics[width=8.0cm]{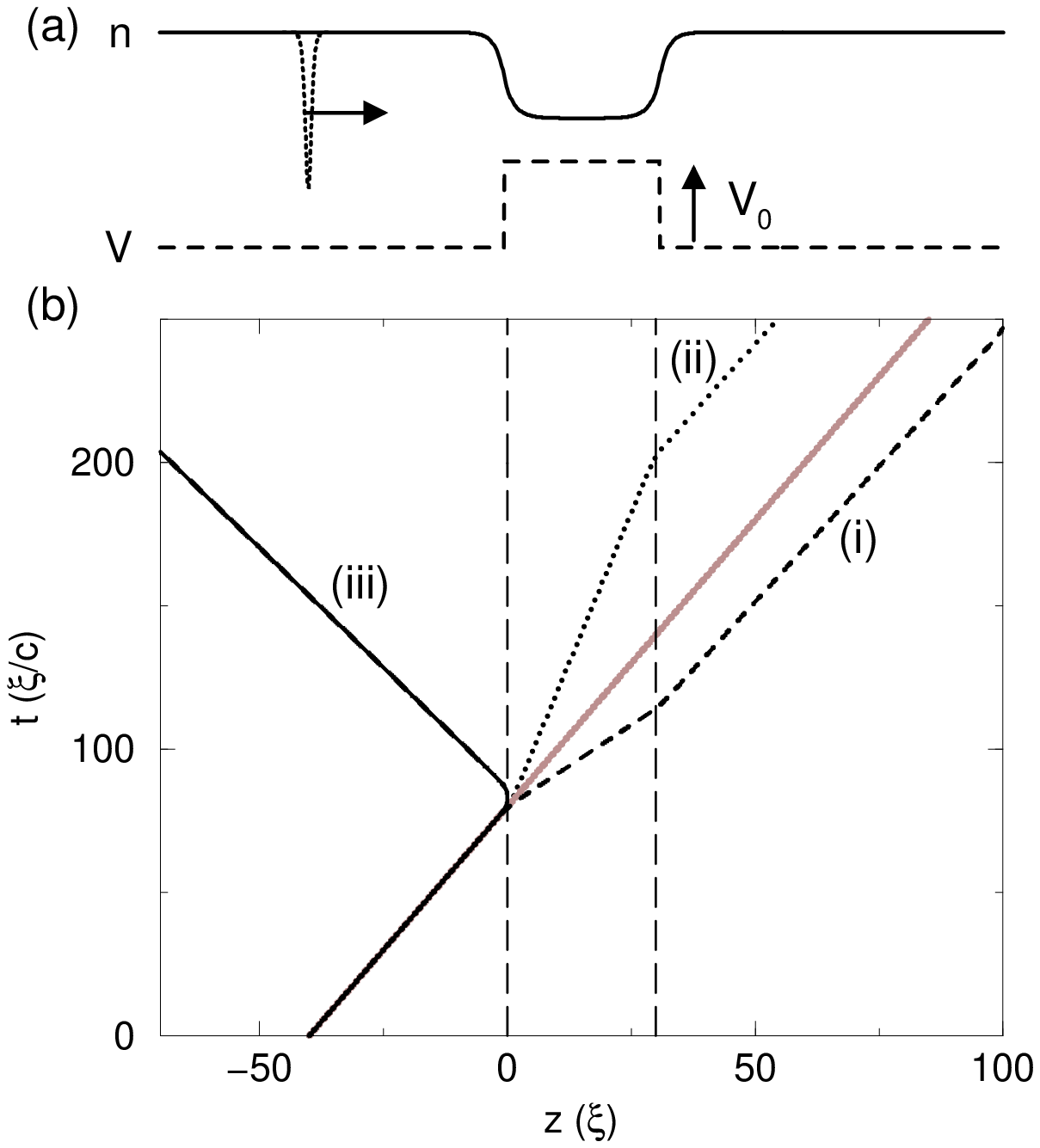} \\
\includegraphics[width=12.0cm]{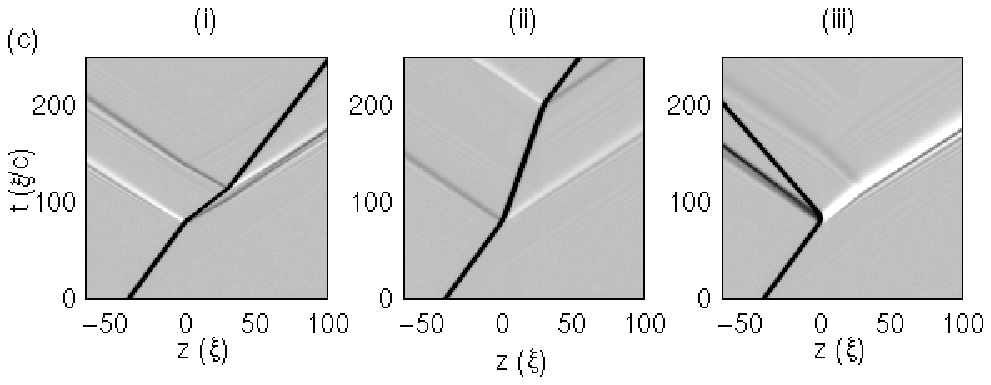}
\caption{(a) Schematic of a dark soliton (dotted line) on a
longitudinally homogeneous condensate background (solid line)
incident on a finite potential step of height $V_{0}$ and length
$30\xi$ (dashed line). (b) Paths of a soliton of initial speed
$v=0.5c$ and position $z=-40\xi$. (i) For $V_{0} < 0$
($V_0=-0.5\mu$, dashed line), the soliton always transmits over
the step; (ii) for positive and small enough $V_0$, the soliton
transmits ($V_{0}=0.2\mu$, dotted line); and, (iii) for large and
positive $V_0$ the soliton is reflected at the first interface
($V_{0} =0.5\mu$, solid black line). For ease of comparison, the
soliton dynamics in the absence of the potential step is also
included here (grey line). (c) Space-time carpet plots of the
renormalised density for cases (i)-(iii) highlight the emission of
counter-propagating sound waves (white/grey lines) whenever the
soliton (black line) interacts with a boundary. These sound waves
typically have an amplitude of a few percent of the peak density.
Note that the emitted sound waves are also refracted at the interface (see, e.g. top left sound wave 
propagation of (c)(i)), albeit in a less pronounced manner, compared to the
solitons.
These and subsequent simulations are performed by the 1D GPE, in which
the chemical potential is set by fixing the peak density to one.
}
\end{center}
\end{figure}

The analogous situation in the case of dark solitons in BECs is
propagation into a region of different potential energy. This
situation can be achieved by the addition of a finite size
potential step along the condensate long axis, leading to a change
in the local background density over a finite length, a situation
illustrated schematically in figure 4(a). For sufficient wide
steps, as employed here, the density in the step region is reduced
to the Thomas-Fermi value, $n(z)=n_0-V(z)$. The behaviour of the
soliton (figure 4(b)) depends crucially on whether the potential
step is positive or negative. In the case of a lower potential
energy region, i.e. higher density (dashed line), the soliton
transmits through the intermediate region. Upon striking each
interface, the soliton emits two counter-propagating sound pulses,
as shown in figure 4(c)(i). A higher density in the step region
leads to a higher speed of sound. If the step height is small and
positive (dotted line in figure 4(b), figure 4(c)(ii)), then the
situation is similar, with the main difference being that the
speed of sound on the step is reduced. However, when the step
height exceeds a critical limit, the soliton is  reflected
(possibly after spending a long time on the interface) (solid line
in figure 4(b), figure 4(c)(iii)).  To first order, the soliton,
with depth $(n_0-v^2)$, will be reflected when the density at
the step becomes too shallow to support it, i.e. when $V_0 > v^2$
(assuming the density at the step heals to the Thomas-Fermi
value).  In actual fact, the soliton becomes reflected at a
slightly higher step height due to the effects of sound emission,
which causes a small decrease in the soliton depth. This effect is
described in detail in \cite{parker2}. Further increases in step
height tend to restrict the sound emission, and in the limit of
collision with a hard wall, $V_0 \gg \mu$, the soliton reflects
elastically.  The loss in the soliton energy, caused by the sound
emission leads to an ultimate increase in the soliton speed. The
difference between initial and final soliton speeds is illustrated
in figure 5, for both cases of higher and lower potential steps.

It should be noted that a qualitatively similar scattering behavior of dark
solitons has also been found using the
equivalent particle theory in the limiting case where the potential step
becomes a localized repulsive impurity \cite{frantzeskakis}.

\begin{figure}
\begin{center}
\includegraphics[width=10.0cm]{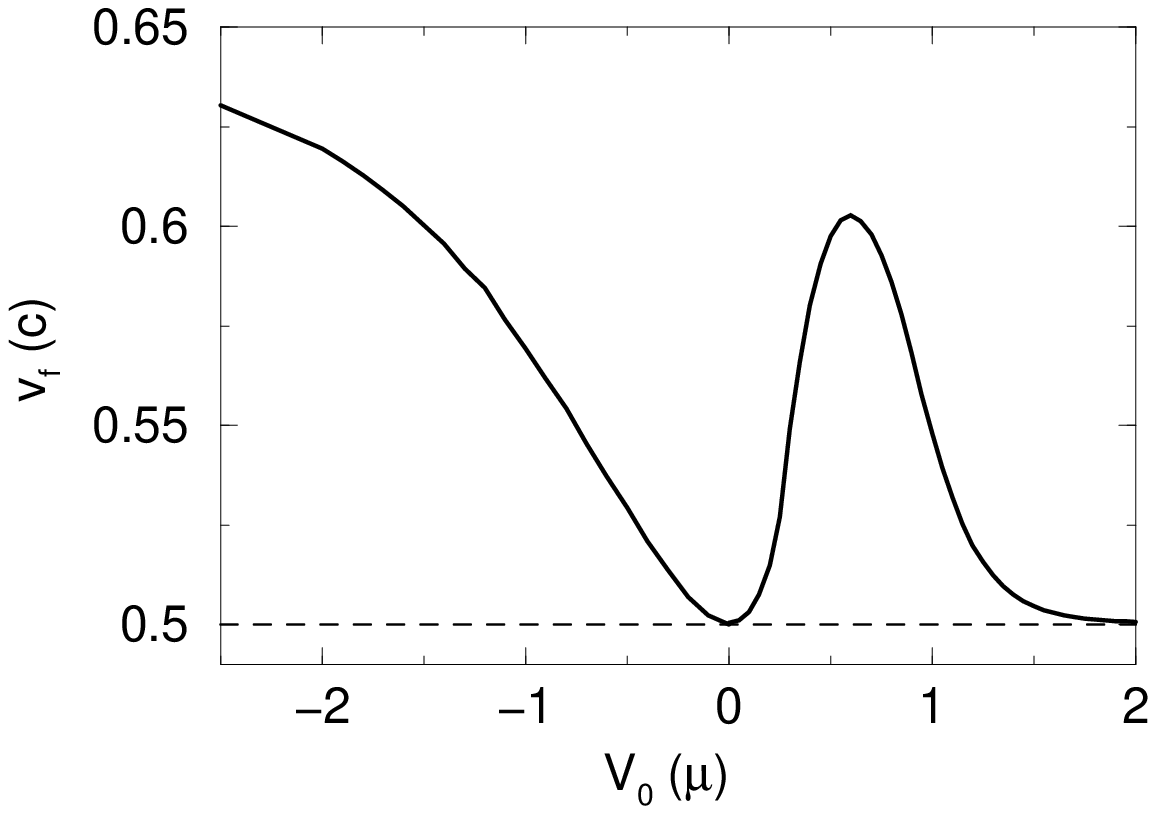}
\caption{Final soliton velocity (solid line) after the interaction of a soliton, with initial speed
$v=0.5c$ (dashed line), with a potential step of height $V_0$
and length $L=30\xi$.  Finite step heights lead to an increase in the soliton speed
due to the process of sound emission. For
$V_0<0$ the soliton always transmits through the step region, with
the emitted sound density, and therefore the final soliton velocity, increasing  monotonically
with $|V_0|$.  Similarly, for $V_0>0$ up to some
critical value, the soliton transmits, and the final velocity
increases. Above this critical point, when the soliton becomes
reflected, further increases of $V_0$ tend to reduce the velocity
change, and for $V_0>2\mu$ the soliton reflects elastically from
the boundary with no change in its speed.}
\end{center}
\end{figure}

\section{Longitudinal Instabilities}

\subsection{Optical Dark Solitons}

As mentioned in the Introduction, optical dark solitons formed in
quasi-1D homogeneous geometries (i.e., temporal ones in optical
fibers or spatial ones in weakly nonlinear bulk media and
dielectric waveguides) are not prone to transverse perturbations
(i.e., to the snake instability). As a result, propagation based
on equation (1) (with $d=1$) will be stable. However, especially
with spatial dark solitons, they are experimentally formed in
media characterized by a strong nonlinearity-induced change of the
refractive index, e.g., alkali vapors (e.g., Rb), semiconductors
(e.g., CdS), photorefractive crystals (e.g., SBN),
photorefractive-photovoltaic crystals (e.g., LiNbO$_{3}$), or
polymers. In such cases, the function $F(I)$ in equation (1) is no
longer a linear function of the light intensity $(I\equiv
|\psi|^{2})$ (i.e., the nonlinearity is of the non-Kerr type) and
other models, such as the competing, saturable, and transiting
nonlinearities are relevant \cite{pelinovsky}. In the framework of
the nonintegrable (even in the absence of the inhomogeneous
potential term) version of equation (1) with a general
nonlinearity, the stability criterion for dark solitons was
derived in \cite{barashenkov} and, at the same time, the
instability-induced dynamics of the dark solitons was investigated
analytically and numerically in detail in \cite{pelinovsky}. In
the latter work, it has been demonstrated that the instability
development is followed by emission of radiation, which propagates
along the continuous-wave (cw) pedestal inducing an effective
dissipation to the dark soliton. Emission of radiation results in
the acceleration of the dark soliton. In fact, these two
quantities are intricately related, and the energy of the
perturbed soliton decays by an acceleration squared law
\cite{pelinovsky},
\begin{eqnarray}
\frac{{\rm d}E_{\rm s}}{{\rm d}t}=-L_{\rm s}(v,n)\left(\frac{{\rm
d}v}{{\rm d}t}\right)^2.\label{eqn:kiv}
\end{eqnarray}
The coefficient is given by
\begin{eqnarray}
L_{\rm s}(v,n)=\frac{c}{c^2-v^2}&~&\left[\frac{2c^2}{n}\left(\frac{\partial N_{\rm s}}{\partial
v}\right)^2\right.
\\
\nonumber
&~&\left.
+2v\left(\frac{\partial
N_{\rm s}}{\partial v}\right) \left(\frac{\partial S_{\rm s}}{\partial
v}\right)
+\frac{n}{2}\left(\frac{\partial S_{\rm s}}{\partial
v}\right)^2 \right].
\end{eqnarray}
and depends on the evolution of the  total phase slip across the moving soliton $S_{\rm s}$, and the
 number of particles displaced by the soliton $N_{\rm s}=\int\left(n-|\psi|^2\right){\rm{d}}z$ during
this unstable motion. The quantities here have been expressed in
terms of rates of change with $v$, which is one convenient way of
formulating the problem, since $v$ is a continuously changing
variable in this problem,  and is directly related to the instability criterion for dark solitons \cite{pelinovsky}.

\subsection{Dark Solitons in Quasi-1D Atomic BEC's}

The above mechanisms of decay do not apply directly to the atomic BEC case. In this case, the GPE
is known to give, at sufficiently low temperatures, an accurate description of the dynamics of
the system. Hence, in this case, decay will arise from other mechanisms, which could include
quantum fluctuations \cite{polish}, thermal damping from the uncondensed (normal) component of the system
(i.e. the thermal cloud confined in the same trap) \cite{muryshev1},
or the effects of the longitudinal potential \cite{parker1,busch}.
Quantum fluctuations are expected to be more pronounced at small soliton speeds (nearly
black solitons), and should not be that important for faster solitons. In addition to this,
the thermal cloud is heavily suppressed at extremely low temperatures (like the ones in which soliton
experiments have so far been performed), suggesting that this mechanism will only become
significant for much higher temperatures. Hence, the longitudinal
confinement is expected to be the key decay mechanism in the limit considered here. This effect has been
considered in detail in \cite{parker1,parker2}.

\begin{figure}
\begin{center}
\includegraphics[width=10.0cm]{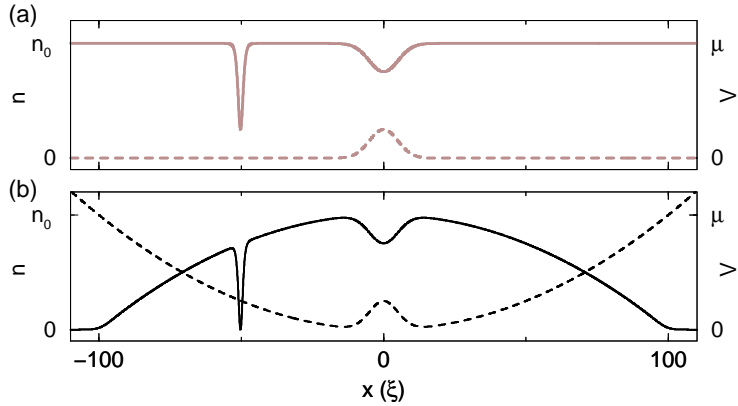}
\caption{A dark soliton with initial position $z=-50\xi$ incident
on a gaussian bump of the form $0.25\exp(-z^2/50)$, for the cases
of (a) homogeneous background density (optical fiber) and initial
soliton speed $v=0.5c$, and (b) longitudinal harmonic
($\omega_z=\sqrt{2}\times 10^{-2} (\mu/\hbar)$ confinement (atomic
traps) and an initially stationary soliton. In each case, the
potential is marked by a solid line, and density by a dashed
line.}
\end{center}
\end{figure}

In this case, the soliton is dynamically unstable through its
entire motion in the system, due to the fact that it constantly
experiences a background density gradient. This is to be
contrasted to the optical case, where instabilities due to modified nonlinearities
 only arise
for a particular range of soliton speeds. The soliton oscillates
in the harmonic trap \cite{rclark,samkeith}, and continuously emits energy in the form of
sound waves \cite{busch}. However, the emitted sound energy remains confined
within the same region, and hence the soliton continuously
re-interacts with the emitted sound field \cite{parker1}.
Nonetheless, one can observe the dissipation of the soliton energy by
either (i) providing a suitable mechanism to damp off the emitted
sound density, or (ii) causing the emitted sound waves to dephase,
with both of these mechanisms leading to experimental proposals
for controlling and measuring the effect of sound emission on
soliton dynamics. In order to damp out the emitted sound, the
soliton can be confined in a tight inner dimple trap, within a
much weaker outer harmonic potential \cite{parker1}. This
situation can be readily realised by focusing an off-resonant
laser beam within a magnetic harmonic trap. If the depth of the
dimple trap is sufficiently shallow, the sound waves can escape to
the outer trap, while the soliton remains confined in this region.
In this limit, sound energy is removed for short enough
timescales, until it bounces off the weaker outer trap and thus
becomes forced to re-interact with the soliton in the inner dimple
region. The second approach relies on soliton motion in a magnetic
trap which is additionally perturbed by an optical lattice
\cite{parker3,theocharis} . In this case, the optical lattice can
confine a soliton within a few lattice sites, with the sound
(again for short enough times) escaping to further located lattice
sites. Although the sound still reinteracts with the soliton, the
presence of the periodic lattice potential dephases the emitted
sound waves, and hence accelerates the decay of the soliton
\cite{parker3}.

\begin{figure}
\begin{center}
\includegraphics[width=10.0cm]{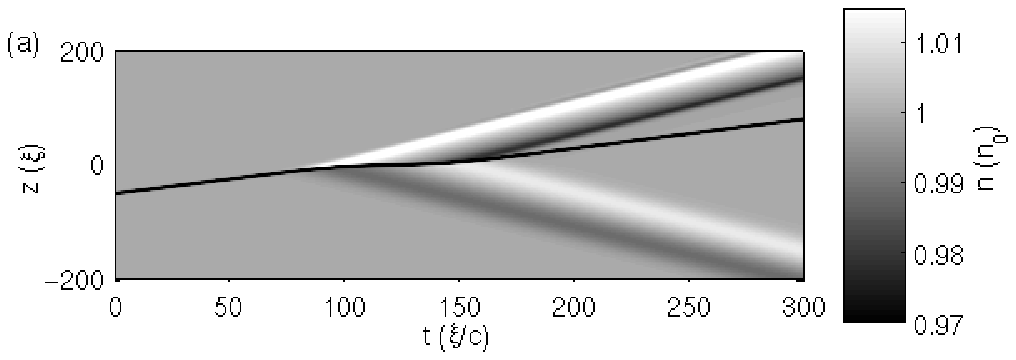} \\
\includegraphics[width=10.0cm]{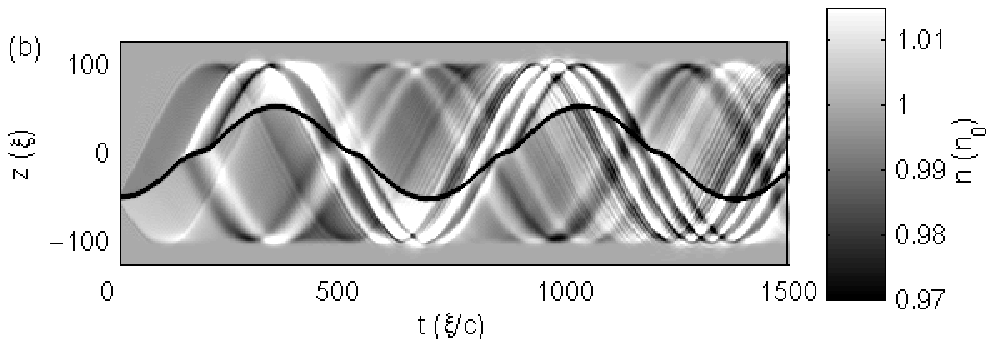}
\caption{Renormalised density space-time plots depicting the long
term evolution of a dark soliton, initially at $z=-50\xi$,
incident on a gaussian bump $0.25\exp(-z^2/50)$ in the cases of
(a) a homogeneous system, and (b) a harmonic trap with
longitudinal trapping $\omega_z=\sqrt{2} \times 10^{-2}$
($\mu/\hbar$). In (a) the soliton, with initial speed $v=0.5c$,
emits counter-propagating sound pulses as it ascends and descends
the bump. (b) The presence of the harmonic confinement induces
periodic oscillations by the initially stationary solition, and
hence repeated crossings over the bump. Although the actual
emission process between these two cases is very similar (see figure 8), the harmonic
potential traps the emitted sound, resulting in continuous
soliton-sound interactions. The presence of the bump additionally
dephases the emitted sound waves, resulting in the ultimate decay
of the soliton after a sufficient number of passes over the bump.
Note the different timescales and lengthscales in the two
pictures.}
\end{center}
\end{figure}

To illustrate the effect of sound emission due to longitudinal
background densities, consider the simple case of a dark soliton
incident on a gaussian bump, as illustrated in figure 6. In the homogeneous case, the soliton
will ascend and descend the bump, and emit two counter-propagating
waves during its interaction with the bump (figure 7(a)). These waves travel off
to infinity and never reinteract with the soliton. The soliton
speed is changed slightly due to the process of sound emission, as
already discussed in section 3. However, the case of a gaussian
bump (typically generated by a focused repulsive blue-detuned
laser beam) in a system with longitudinal harmonic confinement is
drastically different (figure 7(b)). The reason is that, in the absence of the
bump, a dark soliton in a harmonic trap oscillates periodically in
the trap, at a rate approximately equal to $\omega_{\rm
trap}/\sqrt{2}$ \cite{fedichev,parker1,frantzeskakis,busch,huang,lphys,brazhnyi},
where $\omega_{\rm trap}$ is the longitudinal trap frequency. As
the soliton oscillates in the trap, it continuously emits
counter-propagating sound waves, due to the background density
gradient which breaks the integrability of the system. However,
the emitted sound remains confined within the trap, thus
continually re-interacting with the soliton. This leads to a
steady-state in which the soliton is stabilised against decay due
to the complete reabsorption of the emitted sound
\cite{parker1,busch}. The presence of the bump at the centre of
the trap induces further dynamical instability in the soliton,
with the {\it additional} sound emission being similar to that
encountered in the homogeneous system (see figure 8 below). This
additional decay mechanism leads to a more complicated carpet
plot, shown in figure 7(b), and eventually to the gradual decay of
the dark soliton into a sound wave.

Focusing now on the initial part of the motion of the two systems,
i.e. the interaction with the bump, figure 8
 compares the homogeneous limit to harmonically confining traps, highlighting their
similarities and differences. 
The sound emission due to the harmonic confinement means that the
soliton arrives at the trap centre at different times in the trapped system
compared to the homogeneous one. In order to compare the soliton dynamics,
including the interaction of
the soliton with the bump in the two cases, we hence find it convenient to shift
the time axis of figure 8 with respect to that of figure 7, 
such that $t=0$ corresponds to the
time when both dark solitons corresponding to the homogeneous and the trapped 
cases reach the centre
of the bump.
Figure 8(a) shows the background
densities experienced by the soliton in the two cases, with black
corresponding to the harmonic case, and grey to the homogeneous
limit. As the soliton starts ascending or descending the bump, it
experiences a background density gradient which induces it to emit
sound waves. Since the trap is quite shallow, the main
acceleration experienced (solid lines in figure 8(b)) is due to the gaussian bump
and is therefore similar in both cases. This is due to the fact
that the background density gradient (as evident from figure 8(a))
is comparable in this region. The soliton starts decelerating,
with the deceleration tailing off instantaneously to zero when the
soliton is at the gaussian maximum, after which time the soliton
accelerates down the bump. Far from the bump, there will, of
course, be no sound emission occurring in the homogeneous case.
This is, however, not the case for the harmonically confined
system. The continuous harmonic density gradient induces sound
emission during the entire motion of the system in the trap.
Although this effect is instantaneously quite small (at most a few
per cent of the background density), the cumulative effect can be
significant.

The asymmetrically emitted sound contained within the `soliton
region' leads to an apparent deformation of the soliton profile.
This deformation manifests itself as a shift of the soliton centre
of mass $z_{cm}$ from the density minimum $z_{s}$, where the soliton
centre of mass is defined as
\begin{eqnarray}
z_{cm}=\frac{\int_{\rm s} z \left(|\psi|^2-n\right) {\rm
d}z}{\int_{\rm s} \left(|\psi|^2-n\right) {\rm d}z},
\end{eqnarray}
and the `soliton region' ${\rm S}$ is conveniently taken to be
($z_s\pm5\xi)$. This deformation parameter, which is just another
way of parameterising the instantaneously emitted sound density,
is directly proportional to the acceleration, as indicated by the
open squares and right axis of figure 8(b).

\begin{figure}
\begin{center}
\includegraphics[width=11.0cm]{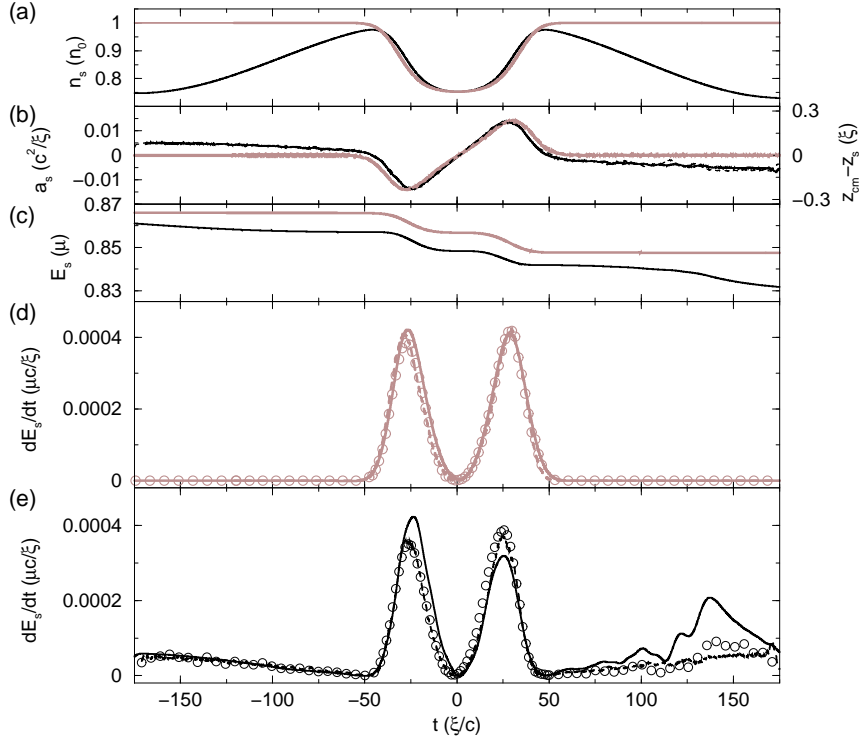}
\caption{ Dynamics of a dark soliton incident on a gaussian bump
$0.25\exp(-z^2/50)$ in the cases of homogeneous (grey lines) and
harmonic (black lines) confinement, corresponding to the cases of
figures 6 and 7. The time axes have been shifted 
such that the soliton reaches the centre of the bump at $t=0$.
(a) Background
density experienced by the moving soliton as a function of time.
(b) Acceleration (solid lines, left axis) and deformation (open squares, right axis) of the
soliton. (c) Soliton energy. (d) Rate of energy loss for the
homogeneous system, as evaluated by the energy functional equation (7)
(solid line), the acceleration squared law of equation (4) (open
circles) and the deformation squared law of equation (8) with a
constant coefficient $\kappa = 0.0078 (\mu c/\xi^3)$ (dashed line)
determined by matching the maximum acceleration and deformation
amplitudes in (b).
 (e)
Same as (d) above, but for the harmonically confined case. The
acceleration squared behaviour breaks down after about $t >
75\xi/c$, when the emitted sound starts reinteracting with the soliton,
after having been reflected off the edge of the trap, leading to
additional sound within the soliton region. }
\end{center}
\end{figure}

The soliton energy is calculated numerically by integrating the
GPE energy functional
\begin{eqnarray}
\varepsilon(\psi)=\frac{1}{2}\left|\nabla\psi\right|^2+V_{\rm
ext}\left|\psi\right|^2+\frac{1}{2}\left|\psi\right|^4,
\end{eqnarray}
across the soliton region S and subtracting the corresponding
contribution from the time-independent background. This procedure
cannot discriminate between soliton and sound energy present in
the interval, and this will be shown to be intimately linked to
the apparent soliton deformation. However, at least in the case of
BECs, one could not discriminate between these two quantities in
the same region.

The soliton energy (figure~8(c)) is intimately related to the
instantaneous acceleration and background density. As such, it
will only decrease when the soliton accelerates, and this occurs
stepwise for the homogeneous system, with instantaneously constant
energy when the system is at the gaussian peak. The emission
process can also be visualised by looking at the rate of energy
loss due to sound emission. The picture in the two cases is again
very similar, and is plotted respectively for homogeneous and
trapped case in figure 8 (d) and (e). The zero emission rate at
the centre of the gaussian bump is due to the locally homogeneous
density and zero acceleration. The maximum magnitude of the sound
emission is found to arise on the sides of the gaussian bump, when
the background density gradient is at a maximum.  The circles
indicate numerical simulations based on equation (4), arising
from nonlinear optics, while the dashed line is based on an
emission law of the form
\begin{eqnarray}
\frac{{\rm d}E_{\rm s}}{{\rm d}t}=-\kappa(z_{cm}-z_{s})^{2}
\end{eqnarray}
with the value of $\kappa$ having been assigned such that it
matches the emission. Although $\kappa$ will in general {\it not}
be a constant, but rather a parameter dependent on the local
soliton speed and background density gradient (similar to the
coefficient of equation (5)), it is remarkable that we still find
such good agreement with a constant coefficient \cite{parker2}. We
have performed detailed quantitative studies of the rate of sound
emission in various geometries \cite{parker1,parker2,parker3}.
Remarkably, we find that this rate is well described by an
acceleration squared law, similar to that of the optical case.
This may at first appear somewhat surprising, given the additional
presence of the inhomogeneous potential in the atomic case, which
was not included in the multiscale perturbative analysis leading
to equation (4) \cite{pelinovsky} . However, in comparing our
results, we have used soliton speeds obtained directly from the
numerical simulations of the GPE, which hence take into account
the modification on soliton speed and phase due to the harmonic
confinement. In principle, one could rederive these equations in
the presence of the harmonic confinement, although we do not
consider this necessary here. To do this, one should combine the
previously employed multiscale asymptotic techniques with boundary
layer theory, as discussed in \cite{busch}.  At a time of $t
\sim 75 (\xi/c)$, we see a deviation of this law. This is due to
the fact that the emitted sound has already travelled to the edge
of the system (i.e. a point in the trapping potential beyond which
it cannot ascend) and become reflected, thus returning towards the
centre of the trap and reinteracting with the soliton. In the
absence of the gaussian bump, such continuous interactions between
soliton and emitted sound actually leads to the stabilisation of
the soliton \cite{parker1}.

\section{Soliton Stabilisation}

In the case of optical dark solitons, inherent linear and
nonlinear losses (the latter is related to two-photon absorption)
of the host medium \cite{optical,kivshar1} affect the cw
background and render the dark soliton unstable. Generally
speaking, losses can be compensated upon introducing a linear
(intensity independent) gain. Nevertheless, although the latter
stabilizes the background, the dark soliton remains unstable
\cite{pertky} and, as a result, dark solitons can be stabilized if
the gain is nonlinear \cite{ikeda}. Other stabilizing techniques
include synchronized phase modulation \cite{kodama}, as well as
phase-sensitive amplification and spectral filtering
\cite{kim}. Additionally, robust parametrically driven dark
solitons, immune from instabilities for all damping and forcing
amplitudes have recently been reported \cite{igor}.

Generally speaking, parametric driving techniques may find application in the context of BECs, as a means to compensate the dissipative losses of dark solitons. In this case, an energy source is needed to compensate the continuous energy loss of the dark solitons due to the various decay mechanisms mentioned above.
 Work towards this direction is currently in progress. Importantly, even if this technique can not be applied in the BEC context, energy can be actually pumped into dark solitons via appropriate perturbations that may be realised by periodic potentials, somewhat analogously to the mechanism by which vortices acquire angular momentum from a laser stirrer. Soliton lifetimes can be noticeably enhanced in this way, and these results are presented elsewhere \cite{new}.

\section{Vortex Dynamics}

Although this paper deals with dark solitons, one should also
mention a few words about other topological structures, such as
vortices. In fact, vortices are not all that distinct from
dark solitons, since, as outlined in section 2, a 3D dark soliton decays
into vortex-like structures. These have already been observed in
both optical and atomic systems.

An optical vortex represents a phase singularity in an optical
beam, and is characterised by a helical wavefront and a dark core
of zero intensity.  The phase of the electric field integrated
around the vortex core is an integer multiple of $2\pi$.  Optical
vortices were first observed in \cite{swartzlander} using a
self-defocussing thermal nonlinearity and a phase mask which
imposed an approximately helical phase structure of the vortex on
the input beam. Another method for imprinting the above mentioned
helical structure, based on a computer generated phase mask, was
suggested in \cite{heck} and implemented in relevant experiments
\cite{ove}. On the other hand, to observe rotating optical
vortices, a similar technique was used  to create a pair of
vortices with the same topological charge \cite{luther}. 
Optical vortices may find applications as ``optical tweezers"
\cite{simpson}, while in a self-defocussing medium, an optical
vortex acts as a dark waveguide \cite{swartzlander}.

The motion of an optical vortex filament due to both the
interaction of other vortices and the presence of the background
electric field has been shown to be strongly analogous to the
hydrodynamic case \cite{rozas1}.  For example, a pair of identical
optical vortex filaments in a gaussian beam have been observed to
rotate about each other at a rate inversely proportional to the
square of their separation \cite{rozas2}.  This result is exactly
analogous to that for corotating vortices in an inviscid
incompressible fluid. However, in a linear medium, the closely
situated vortex cores rapidly expand into each other due to
diffraction effects, while longer rotation angles can be achieved
in self-defocussing media \cite{rozas3}.

In atomic BEC's, quantized vorticity can be excited by phase imprinting \cite{Vortex_Imp,murray}, by
stirring or rotating the system by magneto-optical techniques \cite{vortex_exp1}, or
by the transverse decay of a dark solitary wave \cite{vortex_exp2}, and can appear in
the form of single vortices, vortex lattices \cite{vortex_lattice} and vortex rings. A
single vortex tends to follow a line of constant potential and so
an off-centered vortex will follow a circular trajectory around
the trap centre \cite{anderson2,fetter}. However, vortices are subject to a similar
dynamical dissipation mechanism under acceleration as dark
solitons, leading to the emission of sound \cite{lundh}. If the
emitted sound is suitably damped away the vortex will lose energy
and spiral outwards \cite{vortex_kevrekid,vortex_PRL}.

As an example of the dynamic instability of vortices, we consider
here on the case of a closely positioned co-rotating vortex pair
(two vortices of the same topological charge) in a homogeneous
system. In analogy to the optical case outlined above, the
vortices, separated by $2r$, will rotate about their common axis
at a rate proportional to $1/r^2$, due to their mutual
interaction. It has been shown analytically by Klyatskin
\cite{klyatskin, pismen} that such a co-rotating pair will emit
sound energy at a rate
 proportional to $1/r^6$, and so slowly spiral apart. This result was obtained by looking at the far field
emission pattern, and is hard to verify numerically. Instead, we
illustrate the emission of sound in this process, by solving the
GPE for a homogeneous two-dimensional system. For the co-rotating pair of figure 9(a), the
emitted sound is shown (on much larger scale) in figure 9(b). The
sound emitted by the rotating vortices creates a striking swirling
pattern, which spreads outwards over time.

\begin{figure}
\begin{center}
\includegraphics[width=6.2cm]{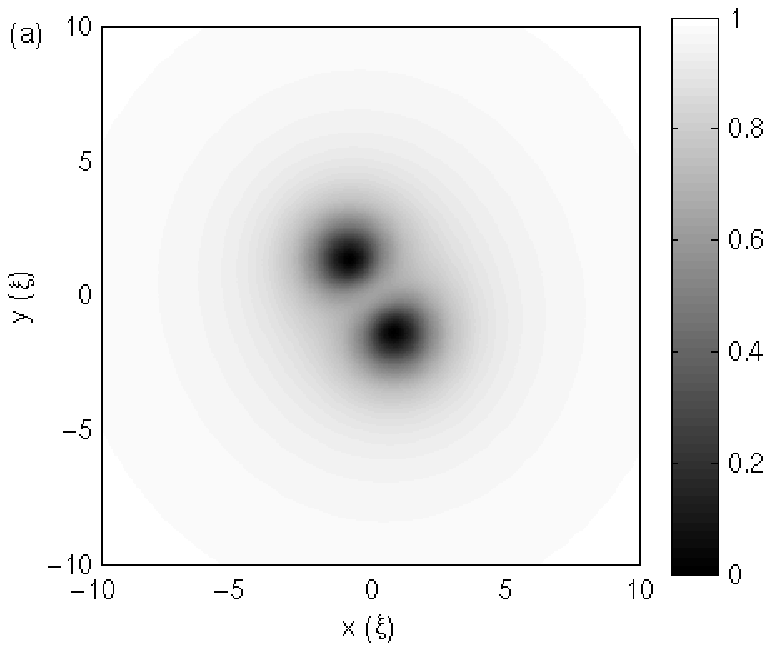}
\includegraphics[width=6.3cm]{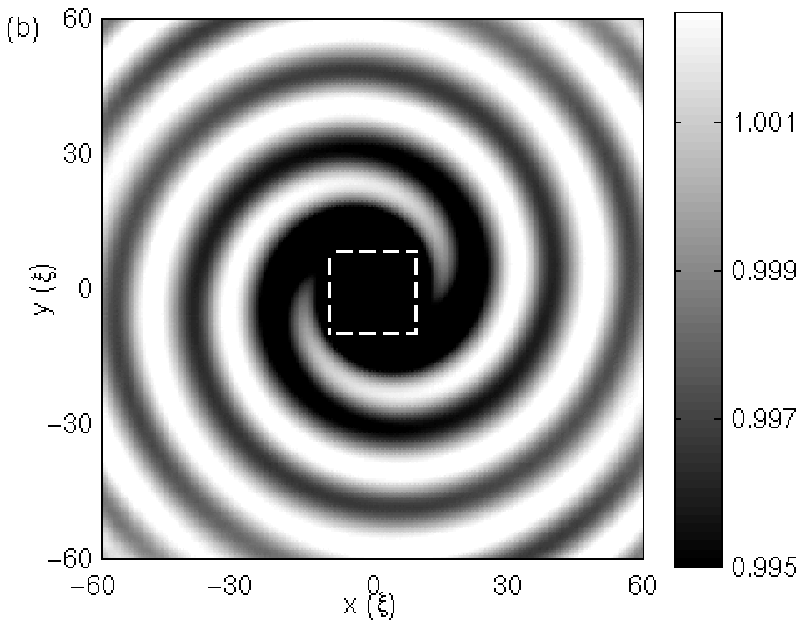}
\caption{ Two-dimensional density plots of a co-rotating vortex
pair in a homogeneous system. The images are plotted on different
greyscale contrast to highlight (a) the vortex pair and (b) the
emitted sound pattern. Note that the sound profile is plotted on a
much larger length-scale, with the region of image (a)
corresponding to the dashed box in the centre of (b). }
\end{center}
\end{figure}

\section{Conclusions}

Many of the well-known nonlinear optics effects can now be readily
observed in assemblies of ultracold Bose-Einstein condensed atomic
gases. In the latter case, the wavefunction of the system obeys,
at temperatures where the thermal component is suppressed, an
equation similar to the equation for the electric field envelope
in cubically nonlinear (Kerr) optical media. The nonlinearity in
the atomic case arises from atomic interactions. This similarity
leads to the observation of common phenomena, such as nonlinear
wave mixing, bright and dark solitons and vortices, with this paper
discussing at length the analogies between dark solitons in these
two very different systems. In this context, we discussed the
snake instability of dark solitons, arising from
excitation of transverse modes in three-dimensional system.
In the limit of tight
transverse confinement, where such mechanisms are heavily
suppressed, we showed that the leading decay mechanisms differ in
the two systems. In real optical media used in experiments, one
actually deals  with a non-Kerr nonlinearity (e.g., a competing or
a saturable nonlinearity), which renders the underlying system
nonintegrable (even in the absence of the potential term, which
may be used to describe a linear graded refractive index change).
As a result, the dark solitons are unstable and their instability
is accompanied by emission of radiation (or sound waves) that
induce an effective dissipation to the dark solitons. In the limit
of very low temperatures (where thermal effects can be ignored),
the main instability in quasi-one-dimensional atomic condensates
actually arises from sound emission along the line of motion, due
to the harmonic confinement experienced (in stark contrast to the
homogeneous optical media with a spatially-independent linear
refractive index).
In the absence of additional
dephasing mechanisms (e.g. the additional presence of a gaussian bump, or a perturbing optical
lattice), the continuous interaction between the oscillating dark
soliton and the co-confined emitted sound actually leads to a steady-state
preventing decay due to {\it longitudinal} confinement. Finally, vortex dynamics in both systems
experience similar effects, with sound emission arising from the
acceleration of vortices, due to either other topological charges,
or background density gradients.

\ack We acknowledge financial support from the UK EPSRC and the Special Research Account of the
University of Athens.


\section*{References}
\begin{thebibliography}{99}

\bibitem{dodd} R.K. Dodd, J.C. Eilbeck, J.D. Gibbon, and H.C. Morris, {\it Solitons and Nonlinear Wave Equations} (Academic Press, London, 1982).

\bibitem{water} J.L. Hammack and H. Segur, J. Fluid Mech. 65, 289 (1974).

\bibitem{macro} M. Peyrard (ed.), Molecular Excitations in Biomolecules
(Springer-Verlag, New York, 1995).

\bibitem{acoustics} K.A. Naugol'nykh and L.A. Ostrovsky, Nonlinear Wave Processes
in Acoustics (Cambridge University, Cambridge, 1998).

\bibitem{plasma} H. Kim, R. Stenzel and A. Wong, Phys. Rev. Lett. 33, 886 (1974);
H-H. Chen and C-S. Liu, Phys. Rev. Lett. 37, 693 (1976).

\bibitem{elastic} A.M. Lomonosov, P. Hess and A.P. Mayer, Phys. Rev. Lett. 88, 076104 (2002).

\bibitem{optical} A. Hasegawa and Y. Kodama, {\it Solitons in Optical Communications}, (Oxford University Press, Oxford, UK, 1995); Yu. S. Kivshar and G.P. Agrawal, {\it Optical Solitons: From Fibers to Photonic Crystals} (Academic Press, Amsterdam, 2003).

\bibitem{exciton} A. Mysyrowitz, Bose-Einstein Condensation, ed. by A. Griffin, D.W. Snoke and S. Stringari (Cambridge University Press, Cambridge, UK, 1995).

\bibitem{dark} J. Denschlag {\it et al}., Science {\bf 287}, 97 (2000). \\
S. Burger {\it et al.}, Phys. Rev. Lett. {\bf 83}, 5198 (1999). \\
Z. Dutton, M. Budde, C. Slowe, and L.V. Hau, Science {\bf 293}, 663 (2001).

\bibitem{bright} K.E. Strecher et al., Nature 417, 150 (2002). \\
L. Khaykovich et al., Science 296, 1290 (2002).

\bibitem{kivshar1} Yu.S. Kivshar and B. Luther-Davies, Phys. Rep. {\bf 278}, 81-197 (1998).

\bibitem{application1} M. Nakazawa and K. Suzuki, Electron. Lett. {\bf 31}, 1076 (1995); {\it ibid} {\bf 31}, 1084 (1995).

\bibitem{application2} B. Luther-Davies and X. Yang, Opt. Lett. {\bf 17}, 1775 (1992); S.Liu {\it et al.}, Appl. Opt. {\bf 36}, 8982 (1997).

\bibitem{application3} P. D. Miller, Phys. Rev. E {\bf 53}, 4137
(1996).
\bibitem{analogies} The relation between atomic and optical solitons has been discussed
in a more mathematical manner in D. Schumayer and B. Apagyi, Phys. Rev. A {\bf 65}, 053614 (2002).
\bibitem{review} F. Dalfovo {\it et al.} Rev. Mod. Phys. {\bf 71}, 463 (1999).

\bibitem{agra} S. Raghavan and G.P. Agrawal, Opt. Commun. {\bf 180}, 377 (2000).

\bibitem{sol} V.E. Zakharov and A.B. Shabat, Zh. Eksp. Teor. Fiz. {\bf 64}, 1627 (1973)
[Sov. Phys. JETP 37, 823 (1973)].
\bibitem{hasegawa} A. Hasegawa and F. Tappert, Appl. Phys. Lett. {\bf 23}, 171 (1973).

\bibitem{emplit} Ph. Emplit {\it et al.}, Opt. Commun. {\bf 62}, 374 (1987).

\bibitem{krokel} D. Kr\"{o}kel {\it et al.}, Phys. Rev. Lett. {\bf 60}, 29 (1988).

\bibitem{weiner} A.M. Weiner {\it et al.}, Phys. Rev. Lett. {\bf 61}, 2445 (1988).

\bibitem{trains} J.A.R. Williams {\it et al.}, Opt. Commun. {\bf 112}, 333 (1994); 
D.J. Richardson {\it et al.}, Electron. Lett. {\bf 30}, 1326 (1994).

\bibitem{swartz} G.A. Swartzlander {\it et al.}, Phys. Rev. Lett. {\bf 66}, 1583 (1991).

\bibitem{allan} G.R. Allan {\it et al.}, Opt. Lett. {\bf 16}, 156 (1991).

\bibitem{duree} G. Duree {\it et al.}, Phys. Rev. Lett. {\bf 74}, 1978 (1995).
\bibitem{Vortex_Imp} M.R. Matthews {\it et al.} Phys. Rev. Lett. {\bf 83}, 2498 (1999).
\bibitem{murray} J.E. Williams and M.J. Holland. Nature {\bf 568}, 401 (1999).
\bibitem{dobrek} L. Dobrek {\it et al.}, Phys. Rev. A {\bf 60}, R3381 (1999).
\bibitem{bongs2} K. Bongs {\it et al.}, J. Opt. B: Quantum Semiclass. Opt. {\bf 5}, S124 (2003).
\bibitem{Dum} R. Dum, J.I. Cirac, M. Lewenstein and P. Zoller, Phys. Rev. Lett. {\bf 80}, 2972 (1998).
\bibitem{carr} L.D. Carr {\it et al.}, Phys. Rev. A {\bf 63}, 051601 (2001).

\bibitem{carr2} S. Burger {\it et al.}, Phys. Rev. A {\bf 65}, 043611 (2002).

\bibitem{gred} S.A. Gredeskul and Yu.S. Kivshar, Phys. Rev. Lett. {\bf 62}, 977 (1989); 
S.A. Gredeskul, Yu.S. Kivshar, and M.V. Yanovskaya, Phys. Rev. A {\bf 41}, 3994 (1990).

\bibitem{gora} D.S. Petrov, G.V. Shlyapnikov and J.T.M. Walraven, Phys. Rev. Lett. 85, 3745 (2000).
\bibitem{lowd} U. Al Khawaja, J.O. Andersen, N.P. Proukakis and H.T.C. Stoof, Phys. Rev. A 66, 013615 (2002).
\bibitem{gorlitz} A. Goerlitz {\it et al.}, Phys. Rev. Lett. {\bf 87}, 130402 (2001).
\bibitem{olshanii} M. Olshanii, Phys. Rev. Lett. 81, 938 (1998).
\bibitem{hansel} W. Haensel et al., Nature {\bf 413}, 498 (2001).
\bibitem{schreck} F. Schreck et al., Phys. Rev. Lett. {\bf 87}, 080403 (2001).
\bibitem{ott} H. Ott et al., Phys. Rev. Lett. {\bf 87}, 230401 (2001).
\bibitem{ed} M.P.A. Jones, C.J. Vale, D. Sahagun, B.V. Hall and E.A. Hinds, Phys. Rev. Lett. {\bf 91}, 080401 (2003).
\bibitem{kavoulakis} A.D. Jackson, G.M. Kavoulakis and C.J. Pethick, Phys. Rev. A {\bf 58}, 2417 (1998).
\bibitem{muryshev} A.E. Muryshev, H.B. van Linden van den Heuvell, and G.V. Shlyapnikov, Phys. Rev. A {\bf 60}, R2665 (1999).

\bibitem{landau} E.M. Lifshitz and L.P. Pitaevskii, Statistical Physics Part 2 (Butterworth, Heinemann, Oxford, 1998).
\bibitem{fedichev} P. O. Fedichev, A. E. Muryshev, and G. V. Shlyapnikov, Phys. Rev. A {\bf 60}, 3220 (1999).
\bibitem{kolomeisky} E.B. Kolomeisky et al., Phys. Rev. Lett. {\bf 85}, 1146 (2000).
\bibitem{polish} J. Dziarmaga, Z. P. Karkuszewski, and K. Sacha,
J. Phys. B {\bf 36}, 1217 (2003).
\bibitem{parker1} N.G. Parker, N.P. Proukakis, M. Leadbeater, and C.S. Adams, Phys. Rev. Lett. {\bf 90}, 220401 (2003).


\bibitem{kuz} E.A. Kuznetsov and S.K. Turitsyn, Zh. Eksp. Teor. Fiz. {\bf 94}, 119 (1988) [Sov. Phys. JETP {\bf 67}, 1583 (1988)]; E.A. Kuznetsov and J.J. Rasmussen, Phys. Rev. E {\bf 51}, 4477 (1995); D.E. Pelinovsky, Yu.A. Stepanyantz and Yu.S. Kivshar, Phys. Rev. E {\bf 51}, 5016 (1995).

\bibitem{ee} D.R. Andersen {\it et al.}, Opt. Lett. {\bf 15}, 783 (1990).

\bibitem{tikh} V. Tikhonenko {\it et al.}, Opt. Lett. {\bf 21}, 1129 (1996);

\bibitem{snake_opt} A.V. Mamaev, M. Saffman and A.A. Zozulya, Phys. Rev. Lett.
{\bf 76}, 2262 (1996); A.V. Mamaev {\it et al.}, Phys. Rev. A {\bf 54}, 870 (1996).

\bibitem{instability} Yu.S. Kivshar and D.E. Pelinovsky, Phys. Rep. {\bf 331}, 117 (2000).

\bibitem{snake_BEC} B.P. Anderson et al., Phys. Rev. Lett. 86, 2926 (2001).



\bibitem{s_th_1} D.L. Feder {\it et al.}, Phys. Rev. A 62, 053606 (2000).

\bibitem{s_th_2} J. Brand and W.P. Reinhardt, Phys. Rev. A 65, 043612 (2002).


\bibitem{kyang} Yu.S. Kivshar and X. Yang, Phys. Rev. E {\bf 50}, R40 (1994).

\bibitem{neshev} A. Dreischuh {\it et al.}, Phys. Rev. E {\bf 66}, 066611 (2002).

\bibitem{gt} G. Theocharis {\it et al.}, Phys. Rev. Lett. {\bf 90}, 120403 (2003).
\bibitem{muryshev1} A. Muryshev {\it et al}, Phys. Rev. Lett. {\bf 89}, 110401 (2002).
\bibitem{komineas} S. Komineas and N. Papanikolaou, Phys. Rev. A {\bf 68}, 043617 (2003).





\bibitem{mn} J.V. Moloney and A.C. Newell, Physica D {\bf 44}, 1 (1990).

\bibitem{aceves} A.B. Aceves, J.V. Moloney and A.C. Newell,
Phys. Lett A {\bf 129}, 231 (1998); J. Opt. Soc. Am. B {\bf 5}, 559 (1988);
Phys. Rev. A {\bf 39}, 1809 (1989); {\it ibid} {\bf 39}, 1828 (1989).


\bibitem{yuriboris} Yu.S. Kivshar and B.A. Malomed, Rev. Mod. Phys. {\bf 61},
763 (1989).

\bibitem{kivshar2} Yu.S. Kivshar, A.M. Kosevich and O.A. Chubykalo,
Phys. Rev. A {\bf 41}, 1677 (1990).

\bibitem{kqt} Yu.S. Kivshar and M.L. Quiroga-Teixeiro,
Phys. Rev. A {\bf 48}, 4750 (1993).

\bibitem{kivshar3} Yu.S. Kivshar, A.M. Kosevich and O.A. Chubykalo,
Phys. Lett. A {\bf 125}, 35 (1987).

\bibitem{kivshar4} Yu.S. Kivshar, A.M. Kosevich and O.A. Chubykalo,
Zh. Eksp. Teor. Fiz. {\bf 93}, 968 (1987) [Sov. Phys. JETP {\bf 66}, 545 (1987)].

\bibitem{parker2}N. G. Parker, N. P. Proukakis, M. Leadbeater, and C. S.
Adams, J. Phys. B {\bf 36}, 2891 (2003).

\bibitem{frantzeskakis} D.J. Frantzeskakis, G. Theocharis, F.K. Diakonos, P. Schmelcher, and Yu.S. Kivshar, Phys. Rev. A {\bf 66}, 053608 (2002).



\bibitem{pelinovsky} D.E. Pelinovsky, Y.S. Kivshar. and V.V. Afanasjev, Phys. Rev. E {\bf 54}, 2015 (1996).
\bibitem{barashenkov} I.V. Barashenkov, Phys. Rev. Lett. {\bf 77}, 1193 (1996).
\bibitem{rclark} W.P. Reinhardt and C.W. Clark, J. Phys. B: At. Mol. Opt. Phys. {\bf 30}, L785 (1997).
\bibitem{samkeith} S.A. Morgan, R.J. Ballagh and K. Burnett, Phys. Rev. A {\bf 55}, 4338 (1997).
\bibitem{busch} T. Busch and J. R. Anglin, Phys. Rev. Lett. {\bf 84}, 2298 (1999).
\bibitem{parker3} N. G. Parker, N. P. Proukakis, C. F. Barenghi
and C. S. Adams, J. Phys. B 37 (In Press, 2004).
\bibitem{theocharis} G. Theocharis, D.J. Frantzeskakis, P.G.
Kevrekidis, R. Carretero-Gonzalez, and B.A. Malomed, subm. to Mathematics and Computers in Simulation (2003).
\bibitem{huang} G. Huang, J. Szeftel, and S. Zhu, Phys. Rev. A {\bf 65}, 053605 (2002).

\bibitem{lphys} N.P. Proukakis, N.G. Parker, C.F. Barenghi and C.S. Adams, Las. Phys. {\bf 14}, 284 (2004).
\bibitem{brazhnyi} V.A. Brazhnyi and V.V. Konotop, Phys. Rev. A {\bf 68}, 043613 (2003).
\bibitem{pertky} Yu.S. Kivshar and X. Yang, Phys. Rev. E {\bf 49}, 1657 (1994).

\bibitem{ikeda} T. Ikeda, M. Matsumoto, and A. Hasegawa, Opt. Lett. {\bf 20}, 1113 (1995); J. Opt. Soc. Am. B {\bf 14}, 136 (1997); X.J. Chen and Z.D. Chen, IEEE J. Quantum Electron. {\bf 34}, 1308 (1998).

\bibitem{kodama} A. Maruta and Y. Kodama, Opt. Lett. {\bf 20}, 1752 (1995).

\bibitem{kim} A.D. Kim, W.L. Kath and C.G. Goedde, Opt. Lett. {\bf 21}, 465 (1996).

\bibitem{igor} I.V. Barashenkov, S.R. Woodford, and E.V. Zemlyanaya, Phys. Rev. Lett. {\bf 90}, 054103 (2003).
\bibitem{new} N.P. Proukakis, N.G. Parker, C.F. Barenghi and C.S. Adams, Preprint (2004).
\bibitem{swartzlander} G. A. Swartzlander, Jr. and C. T. Law, Phys. Rev. Lett.
{\bf 69}, 2503 (1992).
\bibitem{heck} N.R. Heckenberg {\it et al.}, Opt. Lett. {\bf 17}, 221 (1992).

\bibitem{ove} V. Tikhonenko, J. Christou, and B. Luther-Davies, J. Opt. Soc. Am. B {\bf 12}, 2046 (1995); Phys. Rev. Lett. {\bf 76}, 2698 (1996).

\bibitem{luther} B. Luther-Davies, R. Powles, and V. Tikhonenko, Opt. Lett. {\bf 19}, 1816 (1994).
\bibitem{simpson} N. B. Simpson, L. Allen, and M. J. Padgett, J.
Mod. Opt. {\bf 43}, 2485 (1996).
\bibitem{rozas1} D. Rozas, C. T. Law, and G. A. Swartlander, Jr.,
J. Opt. Soc. Am. B {\bf14}, 3054 (1997).
\bibitem{rozas2} D. Rozas, Z. S. Sacks, and G. A. Swartzlander,
Jr., Phys. Rev. Lett. {\bf 79}, 3399 (1997).
\bibitem{rozas3} D. Rozas and G. A. Swartzlander, Jr., Opt. Lett.
{\bf 25}, 126 (2000).
\bibitem{vortex_exp1} C. Raman {\it et al.}, Phys. Rev. Lett. {\bf 83} 2498 (1999).\\
K.W. Madison {\it et al.}, Phys. Rev. Lett. {\bf 84}, 806 (2000).\\
E. Hodby {\it et al.}, Phys. Rev. Lett. {\bf 88}, 010405 (2002).
\bibitem{vortex_exp2} B.P. Anderson {\it et al.}, Phys. Rev. Lett. {\bf 86}, 2926 (2001).
\bibitem{vortex_lattice} J.R. Abo-Shaeer {\it et al.}, Science {\bf 292}, 476 (2001).
\bibitem{anderson2} B.P. Anderson {\it et al.} Phys. Rev. Lett. {\bf 85}, 2857 (2000).
\bibitem{fetter} A. Fetter and A. Svidzinsky, J. Phys. Condens. Matter {\bf 13}, R135 (2001).
\bibitem{lundh} E. Lundh and P. Ao, Phys. Rev. A {\bf 61}, 063612 (2000).
\bibitem{vortex_kevrekid} P.G. Kevrekidis, R. Carretero-Gonzalez, G. Theocharis, D.J. Frantzeskakis and
B.A. Malomed, J. Phys. B: At. Mol. Opt. Phys. {\bf 36}, 3467 (2003).
\bibitem{vortex_PRL} N.G. Parker, N.P. Proukakis, C.F. Barenghi and C.S. Adams, cond-mat/0312520 (2003).
\bibitem{klyatskin} V. I. Klyatskin, Izn. AN. SSSR Mekh. Zh. Gaz
{\bf 6}, 87 (1966).
\bibitem{pismen} L.M. Pismen, {\it Vortices in nonlinear fields}
(Clarendon Press, Oxford, 1999).




\end{thebibliography}
\end{document}